# Generating site saturation mutagenesis libraries and transferring them to broad host-range plasmids using type IIS restriction enzymes

Niels N. Oehlmann[1] and Johannes G. Rebelein[1],[2]*

[1] Max Planck Institute for Terrestrial Microbiology, Karl-von-Frisch-Str. 10, D-35043 Marburg, Germany

[2] Center for Synthetic Microbiology (SYNMIKRO), Philipps-University Marburg, Karl-von-Frisch-Str. 14, D-35043 Marburg, Germany

* Correspondence to johannes.rebelein@mpi-marburg.mpg.de

## Abstract

Protein engineering is an established method for tailoring enzymatic reactivity. A commonly used method is directed evolution, where the mutagenesis and natural selection process is mimicked and accelerated in the laboratory. Here, we describe a reliable method for generating saturation mutagenesis libraries by golden gate cloning in a broad host range plasmid containing the pBBR1 replicon. The applicability is demonstrated by generating a mutant library of the iron nitrogenase gene cluster (*anfHDGK*) of *Rhodobacter capsulatus,* which is subsequently screened for the improved formation of molecular hydrogen.

## 1   Introduction

Site-directed mutagenesis has become one of the most powerful tools in biochemistry and enzymology.[1] The possibility to create large mutant libraries of enzymes and recombinantly produce the enzymes paved the way for the field of enzyme engineering. Screening large libraries allows us to identify enzymes with increased stability, improved activities or new reactivities.[2] For this, the gene of interest is usually introduced and mutated in a suitable plasmid, followed by the transformation of the protein-producing host strain.[3] One strategy for library design is site saturation mutagenesis, where the targeted residue is exchanged with every one of the 20 canonical amino acids.[4] The mutations can be introduced by mismatch primers with the degenerate NNK codon at the position of interest.[5] In cloning-free methods for library construction (e.g., QuickChange methods),[6] the NNK codon has to be located in the primer binding region, which can lead to codon bias during polymerase chain reaction (PCR) amplification. In contrast, the introduction of mutations with scarless cloning techniques such as Golden Gate cloning allows the introduction of mutations via the primer extensions with less influence on the primer binding properties.[7] Golden gate cloning relies on the reactivity of type IIS restriction enzymes that cleave outside their recognition site.[8] Type IIS restriction enzymes enable the introduction of any complementary 4 bp sticky ends at the DNA fragments for seamless ligation.

Plasmids are small circular DNA molecules with a replicon for stable replication and a selection marker (e.g., an antibiotic resistance gene).[9] They are used as cloning vehicles to transform organisms with DNA sequences of interest. The mechanisms of replicons are diverse and can have a limited range of compatible host organisms. Plasmids replicating in a wide range of host organisms are called broad host range plasmids.[10] A member of this group is the pBBR1 plasmid isolated from *Bordetella bronchiseptica* that replicates stably in gram-negative bacteria.[11] Here, we demonstrate the generation of a site saturation mutagenesis library in a broad host range plasmid based on the pBBR1 replicon (Figure 1). The gene of interest, *anfD* (one of the structural genes of the Fe nitrogenase) from *Rhodobacter capsulatus*, is first mutagenized in the *E. coli* cloning plasmid pMM0216 (based on pGGAselect)[12] by PCR with an NNK primer followed by seamless ligation of the amplicon using BsmBI restriction sites introduced by the primer extensions (see Note 1). In a second step, the generated saturation library of *anfD* is transferred into the broad host range plasmid pMM0216 (based on pOG024 created by Philip Poole; Addgene number #113991)[13] by using BsaI restriction sites to insert the mutated *anfD* gene into the *anfHDGK* gene cluster. Using this elaborate protocol, we obtained reproducibly excellent site-saturation libraries in broad host range plasmids with a medium copy number replicon. The library is then conjugated into *Rhodobacter capsulatus* via triparental mating,[14] followed by production and *in vivo* screening of the Fe nitrogenase mutants for hydrogen ($H_2$) formation.

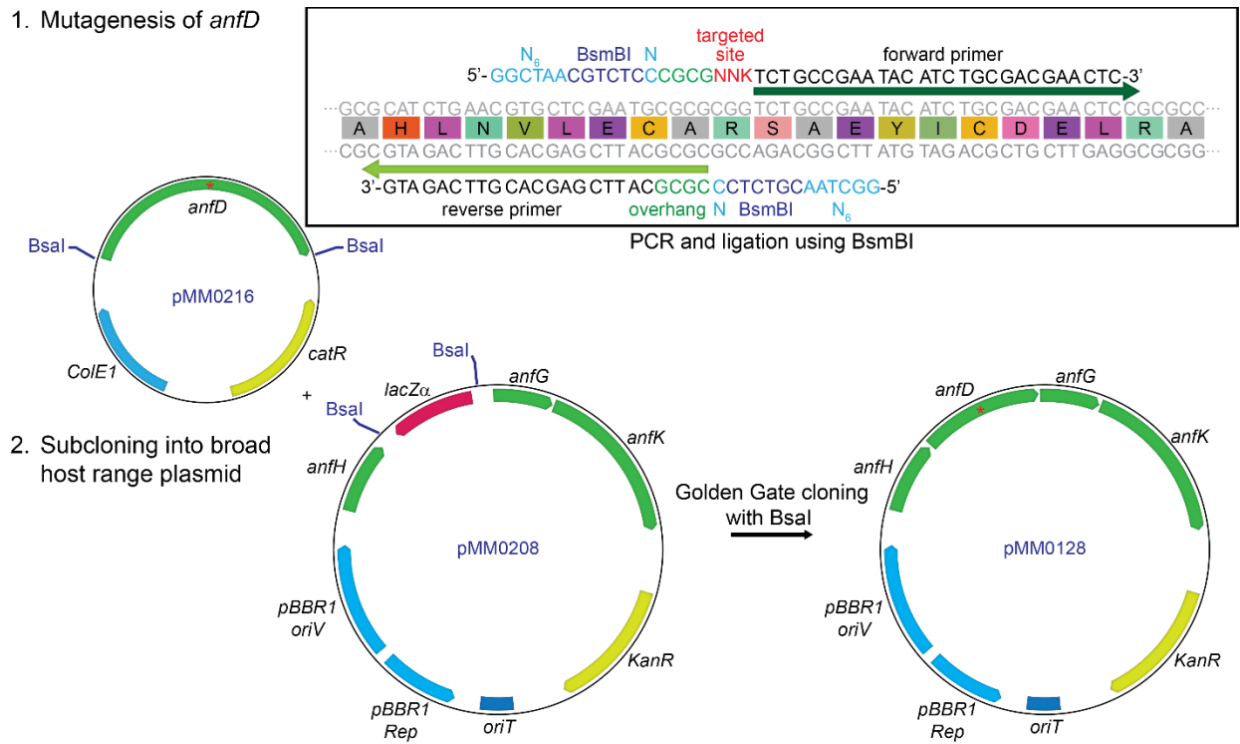

**Fig. 1. Workflow for the saturation mutagenesis library generation of the Fe nitrogenase gene *anfD* in the broad host range plasmid pMM0128.** The *E. coli-specific* high copy number plasmid pMM0216 containing *anfD* is mutagenized by PCR using a mismatched forward primer that introduces the NNK codon at the targeted site. The forward and reverse primer extensions introduce BsmBI cutting sites used to religate the linearized PCR amplicons. Subsequently, the *anfD* library on pMM0216 is cloned into a broad host range plasmid pMM0208 via golden gate cloning using BsaI cutting sites.

## 2   Materials

### 2.1   Bacterial strains, Media and Antibiotics

1. *Escherichia coli*: DH5α is used for DNA cloning, plasmid amplification and transfer to *R. capsulatus* via conjugation. Competent cells of *E. coli* are prepared according to the Inoue method.[15]

2. *R. capsulatus*: MM0167 (Δ*nifD*::*SpR*, Δ*modABC*, Δ*anfHDGK*::*GmR*, Δ*draTG* mutant strain of B10S)[NSMB][16] is used for expression of the *anfD* library in plasmid pMM0128. For general information on the handling and cultivation of *R. capsulatus*, see the more detailed description by Katzke et al.[14]

3. Luria–Bertani (LB) media: 10 g/L tryptone, 5 g/L yeast extract and 10 g/L NaCl dissolved in deionized water

4. RCV minimal medium: 30 mM DL-malic acid, 10 mM serine, 0.8 mM $MgSO_4$, 0.7 mM $CaCl_2$, 50 µM $Na_2EDTA$, 30 µM thiamine·HCl, 45 µM $H_3BO_3$, 9.5 µM $MnSO_4$, 0.85 µM $ZnSO_4$ and 0.15 µM $Cu(NO_3)_2$ with pH adjusted to 6.8. After autoclaving, 10 mM potassium phosphate buffer, 120 µM $FeSO_4$ and 10 mM serine are added from sterile filtered stock solutions. RCV is prepared as described by Katzke et al.[14]

5. PY medium: 10 g/L peptone, 0.5 g/L yeast extract, after autoclaving 2 mM $MgCl_2$, 2 mM $CaCl_2$, 80 µM $FeSO_4$ are added.

6. SOC medium: 2 g/L tryptone, 0.5 g/L yeast extract, 10 mM NaCl, 2.5 mM KCl, 10 mM $MgCl_2$, 10 mM $MgSO_4$ and 20 mM glucose.

7. Agar plates: 1.5% (w/V) agar-agar is added to liquid media before autoclaving.

8. For blue-white screening, prepare a stock solution of 100 mM propan-2-yl 1-thio-*β*-D-galactopyranoside (IPTG) in $ddH_2O$ and 20 mg/ml X-gal in DMSO and store at –20 °C.

9. Autoclave 80% glycerol (V/V) for the generation of cryogenic stocks of libraries and picked clones.

10. Antibiotics: The culture media is supplemented to a final concentration of 50 µg/mL kanamycin (Km, dissolved in $ddH_2O$), 20 µg/ml streptomycin (Sm, dissolved in $ddH_2O$) or 34 µg/ml chloramphenicol (Cm, dissolved in ethanol), as indicated.

2.2 Plasmids

1. pMM0216 (Fig. 1): The cloning vector for *anfD* contains the high copy number ColE1 replicon for stable replication in *E. coli* and the *catR* gene conferring resistance to Cm. A silent point mutation in the codon for *anfD* residue S373 was introduced to remove a BsmBI restriction site. Two BsaI cutting sites flank the *anfD* gene to allow seamless subcloning into the *anfH-GK* gene cluster of pMM0208.

2. pMM0208 (Fig. 1): The *anfD* recipient plasmid pMM0208 contains the *anfH-GK* gene cluster of *R. capsulatus* with a *lacZα* reporter cassette flanked by two BsaI cutting sites as a placeholder for *anfD*, a broad host range replicon (pBBR1 rep) with medium copy number allowing replication in both *E. coli* and *R. capsulatus*, the *kanR* gene conferring resistance to Km and an <u>oriT</u> site for conjugation. A silent point mutation in the codon for *anfK* residue V280 was introduced to remove a BsaI restriction site.

3. pMM0128 (Fig. 1): A broad host range plasmid for the expression of the *anfHDGK* gene cluster.

4. pRK2013: Helper plasmid for conjugation, containing the RK2 transfer genes for conjugation, the ColE1 replicon allowing replication in *E. coli* and the *kanR* gene conferring Km resistance.[17]

2.3 DNA oligonucleotides

The used oligonucleotides are listed in Table 1. Stock solutions of 100 µM are prepared with ddH$_2$O and stored at –20 °C. Working solutions have a concentration of 10 µM oligonucleotide in ddH$_2$O.

**Table 1**. Oligonucleotide used in this method

| name | Sequence (5'-3') | Information |
|---|---|---|
| P1 | GGCTAACGTCTCCCGCGNNKTCTGCCGAATACATCTGCGACGAACTC | Saturation mutagenesis of *anfD* residue R259 |
| P2 | GGCTAACGTCTCCCGCGCATTCGAGCACGTTCAGATG | |

2.4 Chemicals and buffers for agarose gel electrophoresis

1. TAE buffer (1 mM Na$_2$EDTA, 20 mM acetate, 40 mM Tris)
2. Loading dye
3. DNA ladder
4. DNA dye: GelRed
5. Agarose, 1% (w/V) in TAE buffer

2.5 Enzymes

The enzymes used in this protocol were obtained from New England Biolabs (NEB) but may be replaced by enzymes with corresponding properties. They were applied with the provided buffers at the reaction conditions specified in the methods section.

1. Q5 polymerase (2000 u/mL), Q5 polymerase reaction buffer (x5), 10 mM dNTPs
2. DpnI restriction enzyme (20000 u/mL)
3. T4 ligase (400000 u/mL), T4 Ligase buffer (x10)
4. BsmBI restriction enzyme (10000 u/mL)
5. BsaI restriction enzyme (20000 u/mL)

2.6 Commercial Kits

1. Plasmid DNA preparation kit
2. DNA purification kit for PCR reactions

2.7 Devices

1. Standard laboratory equipment for cultivation of *E. coli* and DNA manipulation: Thermoshaker for transformations via heat shock, 37 °C incubator for cultivation on agar plates, benchtop centrifuge, gel electrophoresis chambers, gel imaging system, etc.
2. Thermocycler to perform PCR reactions and incubation of enzymatic reactions.
3. For the anaerobic cultivation of *R. capsulatus* on PY agar plates, airtight anaerobic jars were used with gas packs to deplete atmospheric oxygen. Cells are illuminated by 2 x 3 60 W light bulbs at 30 °C for anaerobic phototrophic growth.
4. Custom-built LED panels (λ: 850 nm, 420 nm and 470 nm, ~60 µmol photons m$^{-2}$ s$^{-1}$) were used to cultivate liquid cultures of *R. capsulatus*.

5. Custom-built vacuum chambers for cultivating *R. capsulatus* in 96-well multitier plates.

6. A vacuum pump connected to a manifold allowing gas exchange of vacuum chambers and sealed gas chromatography (GC) vials

7. Headspace sampling GC for gas analysis of culture headspace.

2.8 Consumables

1. Standard consumables of a molecular and microbiology laboratory: PCR reaction tubes, centrifugation tubes, Petri dishes, inoculation loops, etc.

2. 96-well microtiter plate and aluminum cover foil.

3. 20 mL GC vials, butyl rubber stoppers, aluminum caps to crimp the GC vials.

4. Disposable cannula and 0.2 µm syringe filter to connect GC vials to the vacuum manifold.

5. Gas cylinders of pure argon (Ar) and 8% $CO_2$ in Ar.

# 3 Methods

3.1 Cultivation of strains

1. *E. coli* is cultivated aerobically on LB agar plates at 37 °C. Liquid cultures are grown in 10 ml LB media in 100 ml Erlenmeyer flasks, shaking at 180 rpm at 37 °C.

2. *R. capsulatus* is grown under phototrophic conditions: All cultivations are conducted under anaerobic conditions. To cultivate *R. capsulatus* on PY agar plates, the plates are placed in airtight anaerobic jars and commercially available gas packs are used to deplete the jar atmosphere from oxygen. The transparent jar is illuminated by light bulbs or LED panels at 30 °C. *R. capsulatus* liquid cultures are cultivated in RCV minimal media in sealed, airtight vials or chambers. Using a gas manifold, the gas phase of the vials or chambers is depleted from oxygen by three to five consecutive evacuation-refill cycles with pure Ar (see Note 2). The vials are connected to the manifold via cannulas attached to 0.2 µm syringe filters to remove contaminants from inflowing gas. The cultures are illuminated with LED panels at 30 °C.

3.2 Mutagenesis of *anfD*

1. Forward primer design (see Fig. 1): The forward primer binding region starts downstream of the targeted site and should have a melting temperature ($T_m$) of ~65 °C (calculated according to SantaLucia).[18] The primer extension begins with six nucleotides $N_6$ at the 5'-end, followed by the BsmBI recognition sequence CGTCTC. After a spacer nucleotide N, the 4 bp overhang is introduced, corresponding to the template sequence, followed by the degenerate codon NNK (see Note 3). In this example, site R259 was targeted (see Table 1 for example oligonucleotide).

2. Reverse primer design (see Fig. 1): The biding region of the reverse primer starts upstream of the targeted site and should have a $T_m$ of ~65 °C. The primer extension contains six nucleotides $N_6$ at the 5'-end, followed by the BsmBI recognition sequence CGTCTC and a spacer nucleotide N.

3. For the PCR reaction (total volume 60 µL), add to $ddH_2O$ (38.8 µL) the Q5 reaction buffer (12 µL), 10 mM dNTPs (1.2 µL), reverse and forward primer (3 µL each), template DNA (pMM0216, 4 ng), and Q5 DNA polymerase in a PCR reaction tube.

4. Incubate PCR in a thermocycler using the settings listed in Table 2 for a touchdown PCR.

Table 2: Thermocycler settings for mutagenesis by PCR using mismatched primers

| Description | Cycles | Temperature | Time |
|---|---|---|---|
| Initial denaturation | 1 | 98 °C | 2:00 min |
| Denaturation | | 98 °C | 30 s |
| Annealing | 32 | 68.2 °C (–0.2 °C per cycle) | 30 s |
| Elongation | | 72 °C | 2:30 min |
| Final elongation | 1 | 72 °C | 4 min |
| Storage | 1 | 12 °C | ∞ |

5. Use 5 µl of the PCR reaction mix to prepare a sample for agarose gel electrophoresis (Subheading 3.3). Correct amplification of the vector DNA is verified by the DNA band of the appropriate size (corresponding to ~3.7 kbp; see Note 4).

6. Add DpnI (1 µL) to the PCR reaction mix. Incubate the reaction at 37 °C for 16-24 h.

7. Conduct a DNA purification of the reaction mix using a commercial DNA purification kit (See Note 5). Elute the DNA in ddH$_2$O (25 µL).

8. For the amplicon ligation, prepare a reaction mix (final volume 20 µL) containing T4 ligase buffer (2 µL), T4 Ligase (0.5 µL), BsmBI (0.5 µL) and the PCR amplicon (150 ng) in ddH$_2$O on ice.

9. Incubate the reaction mix in a thermocycler using the settings listed in Table 3.

Table 3: Thermocycler settings for the amplicon religation

| Description | Cycles | Temperature | Time |
|---|---|---|---|
| Digestion | 30 | 42 °C | 5 min |
| ligation | | 16 °C | 5 min |
| Heat inactivation | 1 | 60 °C | 5 min |
| Storage | 1 | 12 °C | ∞ |

10. The reaction mix is used to transform DH5α cells via heat shock: An aliquot (100 µL) of chemocompetent DH5α cells is thawed on ice (15 min). The ligated amplicon (10 µL) is added and the cells are placed on ice (30 min). Next, the cells are heated to 42 °C (30 s) with a thermoshaker and then quickly placed in an ice bath (5 min). The cells are rescued by adding 1 ml SOC media prewarmed to 37 °C and incubation (37 °C, 400 rpm, 1 h). The suspension is centrifuged (6000 xg, 1 min), the supernatant is discarded, the pellet is suspended in the remaining liquid and plated on an LB agar plate supplemented with Cm.

11. After overnight incubation of the LB agar plate, the single colonies are counted to ensure library oversampling (>96 for NNK library). Then, the colonies are spread on the LB plate with an inoculation loop to pool the library. After an additional ~6 h of incubation at 37 °C, the cells are harvested from the plate with an inoculation loop and resuspended in 1 mL LB media. The plasmids are isolated from the suspension using a commercial plasmid DNA preparation kit (0.25 mL of the cell mixture). The rest of the cell suspension is used to prepare a cryogenic stock (add 0.25 mL 80% glycerol, mix and freeze/store at –80 °C)

12. The library generation is verified using Sanger sequencing.

3.3 Agarose gel electrophoresis

1. Prepare a suspension of 1% (w/V) agarose in TAE buffer. Boil the suspension until the agarose has dissolved completely. Cast the gel and add GelRed as indicated by the commercial supplier.

2. Mix the sample with loading dye.

3. Transfer the gel into the electrophoresis chamber with TAE buffer. Load the DNA ladder and samples into the gel chambers.

4. The electrophoresis is conducted at 100 V for 27 min

5. The stained DNA bands are visualized by UV light in a gel imaging system.

3.4 Subcloning of the *anfD* library into pMM0208

1. Prepare a reaction mix (final volume 20 µL) containing T4 ligase buffer (2 µL), T4 Ligase (0.5 µL), BsaI (0.5 µL), pMM0208 (100 ng) and the *anfD* library in pMM0216 in ddH$_2$O on ice.

2. Incubate the reaction mix in a thermocycler using the following settings (Table 4).

   **Table 4:** Thermocycler settings for the subcloning of *anfD* into pMM0208

   | Description | Cycles | Temperature | Time |
   |---|---|---|---|
   | Digestion | 30 | 37 °C | 5 min |
   | ligation |  | 16 °C | 5 min |
   | Heat inactivation | 1 | 60 °C | 5 min |
   | Storage | 1 | 12 °C | ∞ |

3. The reaction mix is used to transform DH5α cells via heat shock (see Subheading 3.2). The outgrowth is plated on an LB agar plate supplemented with Km, 20 µg/ml X-gal and 100 µM IPTG for blue-white screening.

4. After overnight incubation of the plate at 37 °C, successful integration of the *anfD* library is assessed by the number of white and blue colonies (successful cloning results in hundreds of white and few blue colonies). The single colonies are spread on the plate using an inoculation loop to pool the library, the plate is incubated further for 6 h at 37 °C and the cells are harvested and suspended in 1 ml LB media. The plasmid is isolated from the suspension (0.25 mL) using a commercial plasmid DNA preparation. The rest of the cell suspension is used to prepare a cryogenic stock (add 0.25 mL 80% glycerol, mix and freeze/store at –80 °C).

5. Successful transfer of the *anfD* library is verified by Sanger sequencing

3.5 Conjugation into *R. capsulatus*

1. Two days prior to the conjugation, cultivate *R. capsulatus* strain MM0167 on a PY agar plate supplemented with Sm. Incubate the plate under phototrophic conditions (see Subheading 3.1) at 30 °C.

2. The day before the conjugation, inoculate the DH5α strain carrying the conjugation helper plasmid pRK2013 and the DH5α strain carrying the *anfD* library on pMM0128 on a separate LB agar plate supplemented with Km. Incubate the plates overnight at 37 °C.

3. For the conjugation, harvest the *R. capsulatus* cells with an inoculation loop and suspend them in 1 ml RCV media. In 0.5 mL PY media, suspend an equal amount (~1 inoculation loop) of the *E. coli* DH5α strains carrying pRK2013 or the pMM0128 *anfD* library and add the suspension to the *R. capsulatus* cells. Centrifuge the cell suspension at 13,000 xg for 10 min.

4. Meanwhile, place an autoclaved cellulose acetate filter (see Note 6) on an LB plate (without antibiotics).

5. Decant the supernatant of the cell suspension and suspend the cell pellet in the remaining liquid (~200 µL) by gentle agitation using a pipette tip. Transfer the mixture to the cellulose acetate filter on the LB plate. Transfer the plate carefully to 30 °C incubator and incubate it for 24 h.

6. Harvest the cells from the cellulose acetate filter. This can be done by transferring the filter into a 2 mL centrifuge tube filled with 1 mL RCV media and scraping the cell lawn into the media using a pipette tip. Remove the filter and suspend the cells using a pipette.

7. Plate out 100 µl of the cell suspension and 100 µL of a 1:10 suspension dilution in RCV media onto PY agar plates supplemented with Km and Sm.

8. Incubate the plates under phototrophic growth conditions until single colonies appear (~2 d).

3.6   *In vivo* screen of the *anfD* library for $H_2$ formation

As a proof of concept, a saturation mutagenesis library for *anfD* residue R259 was prepared according to the method described (Subheadings 3.2 and 3.4). After conjugation of the library from *E. coli* strain DH5a into *R. capsulatus* strain MM0167 by triparental mating (described in Subheading 3.5), the library is investigated for the formation of $H_2$ *in vivo* according to the following procedure (Fig. 2).

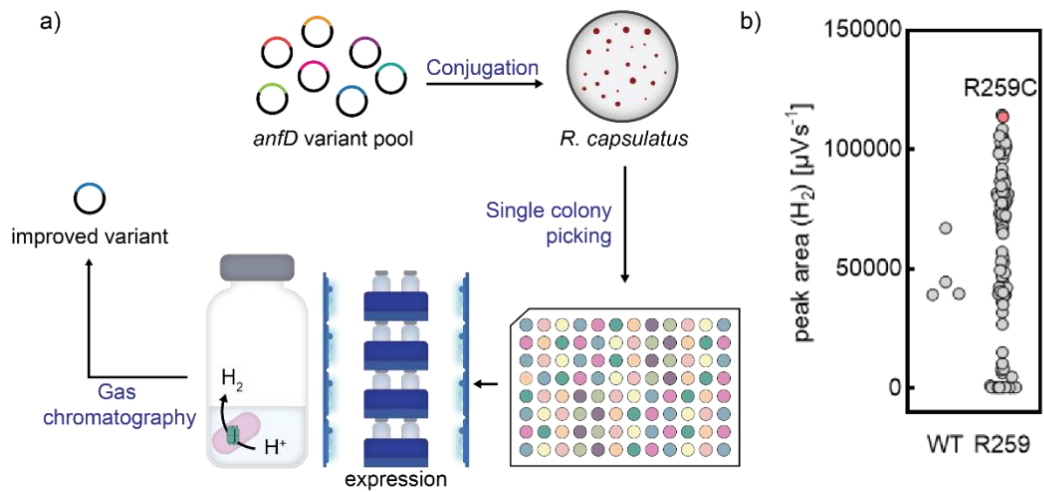

**Fig.2. (a) Workflow of the *in vivo* screening of *anfD* library in *R. capsulatus* strain MM0167 for the formation of $H_2$. (b) Result for the saturation mutagenesis library of *anfD* residue R259.** (a) The variant pool of *anfD* in pMM0128 is conjugated by triparental mating into *R. capsulatus* strain MM0167. Single colonies of the conjugant *R. capsulatus* strain are picked into RCV media in a 96-well plate and cultivated phototrophically. Subsequently, the cultures in the microtiter plate are used to inoculate assay cultures in 20 ml GC vials. The assay cultures are cultivated phototrophically, which results in the recombinant expression of the *anfDGK* gene cluster on pMM0128 and the formation of $H_2$ by the Fe nitrogenase variants *in vivo*. Headspace gas analysis is used to determine the variant with the highest *in vivo* $H_2$ forming capabilities under the applied conditions. (b) The $H_2$ peak area of the headspace gas analysis for the R259 site saturation mutagenesis library and of strains carrying the wild type (WT) pMM0128 sequence are shown after 6 d of phototrophic growth.

1. Fill 250 µL of RCV media supplemented with Km into each well of a transparent 96-well microtiter plate.

2. Pick single colonies of *R. capsulatus* strain MM0167 carrying pMM0128 with the *anfD* library from the conjugation plate into the individual well of the 96-well microtiter plate (e.g., by using small pipette tips).

3. The 96-well plate is placed into a vacuum chamber. To minimize the edge effect (often-observed growth differences of cells cultured in the outer wells of a microtiter plate)[19] during the phototrophic growth of *R. capsulatus*, place a 96-well microtiter plate filled with 250 µL dd$H_2$O in each well underneath the culture plate as a humidity reservoir.

4. The vacuum chamber is sealed airtight and the gas in the chamber is replaced with Ar by four evacuation-refill cycles using a gas manifold. In a final evacuation-refill cycle, the gas in the chamber is exchanged to 8% $CO_2$ in Ar to a final pressure of 1.2 atm.

5. The inoculated 96-well microtiter plate is incubated under phototrophic conditions at 30 °C for 5 d.

6. In 20 mL GC vials, fill 3.8 mL RCV media supplemented with Km. Open the anaerobic chamber and inoculate 200 µL of *R. capsulatus* cell cultures from each well of the 96-well plate into a GC vial. Close the vials with butyl rubber stoppers and crimp the vials airtight. Connect the GC vials with cannulas to a gas manifold and exchange the headspace to 8% $CO_2$ in Ar (see Subheading 3.1).

7. Add 75 µL 80% glycerol to the remaining liquid in each well of the 96-well plate to generate a cryogenic stock, and seal the plate with aluminum seal foil. Mix the plate and store at –20 °C.

8. The 96 vials are incubated under phototrophic conditions for 6 d. The *in vivo* assay is terminated by removing the vials from the light source.

9. A headspace sampling GC-FID/TCD is used to quantify the amount of $H_2$ in the culture headspace of each vial (See Note 7).

10. After the analysis, the vials of interest are opened and the plasmid variant with improved activity is isolated from the cultures using a plasmid DNA preparation kit (See Note 8).

11. The sequence of the plasmid variant is determined by Sanger sequencing

Following the described procedure, we identified a mutation of R259 with significantly improved $H_2$ formation *in vivo* compared to the wild-type *anfD* sequence. The two culture vials with the highest amount of $H_2$ in the headspace expressed the R259C *anfD* mutant. The corresponding mutation R277C in the Mo nitrogenase of the model organism *Azotobacter vinelandii* has been described before and was found to still produce $H_2$ with a higher electron flux distribution towards $H_2$ under an $N_2$ atmosphere compared to the Mo nitrogenase wild-type sequence (WT: 70% $NH_3$, 30% $H_2$; R277C: 49% $NH_3$, 51% $H_2$).[20] In the case of the Fe nitrogenase, a 2.4-fold increase of $H_2$ was detected in the vial headspace of the cultures expressing the R259C mutant compared to the WT *anfD* sequence.

# 4    Notes

1. In our experience, the direct mutagenesis of the broad-host range plasmid pMM0128 (containing the *anfHDGK* gene cluster in plasmids based on the pBBR1 replicon) showed low efficiencies and lacked reproducibility. We solved this issue by introducing the mutations in pMM0216 (containing *anfD* in the well-established pGGAselect plasmid backbone), which is smaller and has a higher copy number than pMM0128.

2. *R. capsulatus* can grow aerobically using $O_2$ as a terminal electron acceptor during respiration. For the depletion of residual dissolved $O_2$ in culture media, it is usually sufficient to start the cultivation and let *R. capsulatus* consume the remaining $O_2$. For large culture volumes or, if necessary, the culture media can be degassed from $O_2$ by purging the media with Ar under vigorous stirring. The gas inlet needle should be connected to a 0.2 µm syringe filter to prevent contamination.

3. If codon bias is observed in the constructed libraries, different commercial suppliers for oligonucleotide synthesis should be tested.  Depending on the commercial supplier, we experienced varying quality/bias of the NNK codons.

4. The appearance of strong unspecific amplification bands caused a drastic drop in efficiency for the following amplicon ligation. Optimizing the PCR conditions or purifying the correct DNA band from an agarose gel could help in such cases.

5. We observed an increase in amplicon ligation efficiency if the amplicon was purified using more washing buffer as described by the protocol of the commercial DNA purification kit supplier.

6. The cellulose acetate filters can be autoclaved but need a small weight on top (e.g., a small glass beaker) to keep their shape. The filters are placed on the LB agar plate with tweezers sterilized by ethanol.

7. This method description focuses on generating mutant libraries in a broad host range plasmid and their expression in *R. capsulatus.* A detailed description of the gas analysis can be found in Schmidt *et al.*[21]

8. Plasmids can be purified from *R. capsulatus* using a plasmid DNA preparation kit according to the protocol for *E. coli* cultures without adaptations.

## Acknowledgment

J.G.R. thanks the Deutsche Foschungsgemeinschaft (DFG, German Research Foundation) – 446841743 for funding. N.N.O. thanks the Fonds der Chemischen Industrie for a Kekulé fellowship. We thank B. Masepohl and T. Drepper for providing strains and plasmids.